\documentclass[aps,prd,preprintnumbers,12pt]{revtex4-1}
\usepackage{amsmath, mathrsfs, amssymb,amsfonts,amsthm,graphicx, epsf, dcolumn, yfonts}
\usepackage[hyperfootnotes=true]{hyperref}
\usepackage{tikz}
\usepackage{color}
\usepackage{slashed}
\usepackage{setspace}
\usepackage{cancel}
\usepackage{wasysym}
\usepackage[lofdepth,lotdepth]{subfig}
\usepackage{float}
\usepackage[normalem]{ulem}
\usepackage[symbol]{footmisc}
\pdfoutput=1
\usepackage[utf8]{inputenc}
\usepackage{ragged2e}
\DeclareCaptionJustification{justified}{\justifying}
 \newcommand{\be}{\begin{equation}}
 \newcommand{\ee}{\end{equation}}
 \newcommand{\bea}{\begin{eqnarray}}
 \newcommand{\eea}{\end{eqnarray}}

\newcommand{\beq}{\begin{equation}}
\newcommand{\eeq}{\end{equation}}

\renewcommand*{\thefootnote}{\fnsymbol{footnote}}

\begin{document}

\thispagestyle{empty}

\title{Computing spacetime}
\author{Juan F. Pedraza,$^{\dag,a}$ Andrea Russo,$^{\ddag,b}$ Andrew Svesko$^{\ddag,c}$ and Zachary Weller-Davies$^{\ast,d}$}

\affiliation{$^\dag$Instituto de F\'isica Te\'orica UAM/CSIC, Madrid, 28049, Spain
\vspace{-0.1cm}\\\vspace{-0.1cm}
$^\ddag$Department of Physics and Astronomy, University College London, London, WC1E 6BT, UK
\vspace{-0.1cm}\\
$^\ast$Perimeter Institute for Theoretical Physics, Waterloo, ON N2L 2Y5, Canada}
\begin{abstract}\vspace{-2mm}
\noindent Inspired by the universality of computation, we advocate for a principle of \emph{spacetime complexity}, where gravity arises as a consequence of spacetime optimizing the computational cost of its own quantum dynamics. This principle is explicitly realized in the context of the Anti-de Sitter/Conformal Field Theory correspondence, where complexity is naturally understood in terms of state preparation via Euclidean path integrals, 
and Einstein's equations emerge from the laws of quantum complexity.
We visualize spacetime complexity using Lorentzian threads which, conceptually,  represent the operations needed to prepare a quantum state in a tensor network discretizing spacetime. Thus, spacetime itself evolves via optimized computation.

\vspace{5cm}

\begin{center}
\emph{This essay was submitted to the 2022 Essay Competition of the Gravity Research Foundation}\\
Submission date: March 30, 2022\\
\end{center}
$\,$\\\vspace{-2mm}
\hspace{-8mm}$^a$\verb"j.pedraza@csic.es"\\\vspace{-2mm}
\hspace{-8mm}$^b$\verb"andrea.russo.19@ucl.ac.uk"\\\vspace{-2mm}
\hspace{-8mm}$^c$\verb"a.svesko@ucl.ac.uk" \\\vspace{-2mm} 
\hspace{-8mm}$^d$\verb"zwellerdavies@pitp.ca" \verb"(corresponding author)"

\end{abstract}

\renewcommand*{\thefootnote}{\arabic{footnote}}
\setcounter{footnote}{0}

\maketitle


\newpage

\setcounter{page}{1}

An important theme in computer science is optimization. That is, the aim to develop an algorithm which performs a task or operation in the most efficient way possible given some set of resources; cost effective computation. Program optimization is ubiquitous to  the extent that entire fields of science, such as biology, may be reformulated in terms of computation. 
In particular, the notion of optimization appears in a fundamental concept of classical physics: the principle of least action. Roughly, the principle of least action says that the optimal path which evolves a system from its initial to final configuration is a solution to the equations of motion, the trajectory for which the action is stationary. In other words, the equations of motion reduce the cost of computing the dynamics of the system. For example, as first postulated by Fermat, light travels between two points along the path of least time. Thus, in an almost teleological fashion, light computes which trajectory minimizes the travel time and chooses its path among an infinite number of alternatives. Applied more broadly, we may rephrase a maxim historically attributed to Maupertuis: Nature is thrifty in its computation.
In this essay we explicitly demonstrate how gravitational dynamics, encoded in Einstein's equations, emerge as a result of spacetime optimizing its computation. 

First, it is natural to frame program optimization in terms of computational complexity. Given an initial reference state $|\psi_{i}\rangle$, and a finite set of gates (represented by unitary operations) $\{g_{1},...,g_{N}\}$, the (quantum) computational  complexity $\mathcal{C}(|\psi_{f}\rangle)$ of preparing a specific target state $|\psi_{f}\rangle$ is equal to the minimum number of such gates needed to construct the unitary operator $U_{fi}$ which transforms $|\psi_{i}\rangle$ into $|\psi_{f}\rangle$,
\beq |\psi_{f}\rangle=U_{fi}|\psi_{i}\rangle=g_{j_{n}}...g_{j_{2}}g_{j_{1}}|\psi_{i}\rangle\;.\label{eq:defcomp}\eeq
Thus, the complexity $\mathcal{C}$ defines the optimal cost required to prepare a specific target state given some initial reference state, within some accuracy. The computational complexity is often interpreted as circuit complexity, a construction in which the reference and target states, together with the set of unitary operations, define a quantum circuit.
More generally, the complexity of preparing a specific target state $|\psi_{f}\rangle$ from $|\psi_{i}\rangle$ can be understood by associating a cost to each mapping $U_{fi}$ given a set of resources, and finding the optimal one.

Extending the notion of computational complexity to quantum field theories is an active area of research, in which there exist multiple working definitions of field theory complexity. One approach is to generalize Nielsen's `geometrization' of circuit complexity \cite{Nielsen:2006,Nielsen:2007} to field theories. In this context, quantum circuits are represented by geodesics in an auxiliary manifold of unitary operations. The length of the minimal geodesic connecting the reference and target states characterizes the complexity, analogous to Fermat's principle of least time. In other words, minimizing computational cost is equivalent to finding minimal length geodesics, such that the optimal program is interpreted as a `free fall' trajectory through a complexity geometry. Phrased like this, it is tempting to reinterpret ordinary free falling motion in the language of complexity and computation. We aim to apply this interpretation to spacetime iteself. 
The dynamics of spacetime are governed by Einstein's equations and are traditionally derived using the principle of least action. That is, in an auxiliary space of spacetime metrics, the minimal `geodesic' yields the metric solving Einstein's equations. Here we show how they emerge from a new principle we term \emph{spacetime complexity}: 

\vspace{2mm}

\noindent \emph{In a quantum theory of gravity, Einstein's equations arise in the (semi)classical limit as a result of spacetime minimizing the cost of computing its own quantum dynamics}.

\vspace{2mm}

To explicitly realize this principle, we work in the context of the Anti-de Sitter/Conformal Field Theory (AdS/CFT) correspondence, where gravity in (`bulk') asymptotically AdS spacetime has a dual description in terms of a  holographic CFT with a large number of degrees of freedom living on the boundary of AdS. 
 In this picture, specific CFT states describe particular asymptotically AdS geometries; for example, the CFT vacuum provides a dual description of empty (vacuum) AdS, and vice versa. More generally, bulk Lorentzian spacetimes describe the time evolution of coherent holographic CFT  states  prepared by Euclidean path integrals with sources turned on \cite{Skenderis:2008dh,Skenderis:2008dg,Botta-Cantcheff:2015sav,Marolf:2017kvq,Botta-Cantcheff:2019apr}.

To prepare a CFT state, one performs a Euclidean path integral over the Euclidean geometry where the CFT is defined, namely, a southern hemisphere.
Mathematically, we prepare a coherent state 
$|\lambda_f\rangle$ from a reference state $|\lambda_i\rangle$ by evaluating a path integral with sources turned on. Heuristically, 
\beq |\lambda_f\rangle = U_{fi}(\lambda_{\alpha})|\lambda_i\rangle= e^{-\int_{\tau<0}d\tau d\vec{x}\sum_{\alpha}\lambda_{\alpha}\mathcal{O}_{\alpha}}|\lambda_{i}\rangle\;,\label{eq:statelambdatest}\eeq 
where $\tau$ is a Euclidean time, with $\tau<0$ representing the southern hemisphere, and $\{\lambda_{\alpha}\}$ denote sources for CFT primary operators $\mathcal{O}_{\alpha}$. The reference state wavefunctional $|\lambda_i\rangle$ itself is likewise represented by a Euclidean path integral. For example, when it is taken to be the  CFT vacuum, the sources sources $\lambda_{i}$ are turned off and $|\lambda_i\rangle=|0\rangle\equiv\int_{\tau<0}[D\phi]e^{-I_{E}^{\text{CFT}}}$, where $I_{E}^{\text{CFT}}$ is the CFT Euclidean action.

\begin{figure}[t]
$\qquad\qquad\quad$\includegraphics[scale=0.45,trim=0 2cm 10cm 12cm]{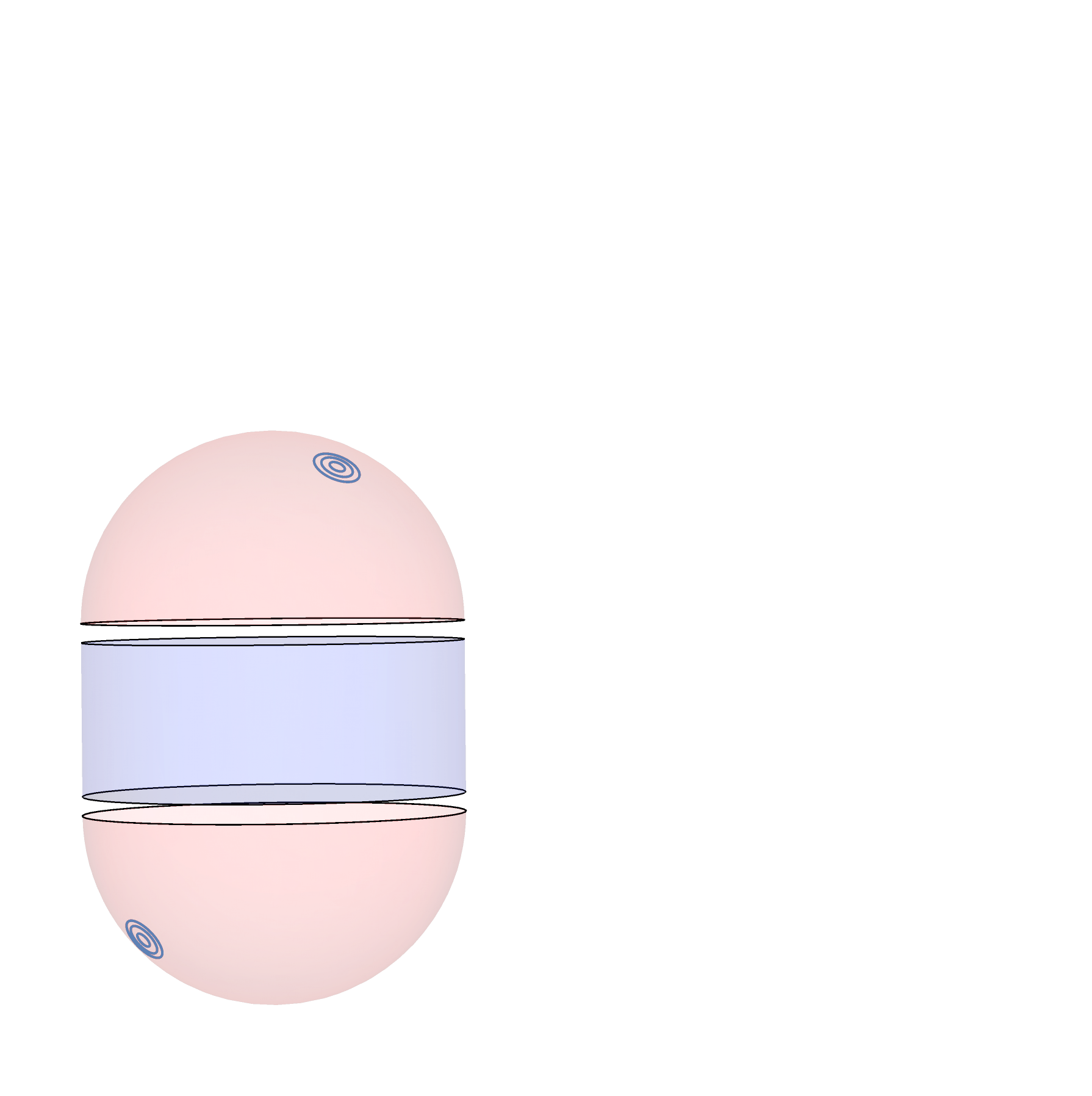}
\centering
\begin{picture}(0,0)
\put(-222,19){$\lambda_f$}
\put(-175,23){$\mathcal{M}_{-}$}
\put(-143,203){$\lambda_f'$}
\put(-120,62){$\Sigma_{-}$}
\put(-120,86){$\Sigma_{-}$}
\put(-175,100){$\tilde{\mathcal{M}}$}
\put(-225,123){$\Sigma_{+}$}
\put(-225,147){$\Sigma_{+}$}
\put(-178,171){$\mathcal{M}_{+}$}
\put(-96,75){$|\lambda_f\rangle$}
\put(-96,138){$\langle\lambda'_f|$}
\end{picture}
\caption{Visualization of state preparation of holographic coherent CFT states using Euclidean path integrals.
The sources and the reference state defined on the southern hemisphere of $\mathcal{M}_{-}$ prepare the target state on $\Sigma_{-}$. Given initial analytic data on $\Sigma_{-}$, Einstein's equations describe Lorentzian evolution in $\tilde{\mathcal{M}}$.
A complete transition amplitude requires one to glue another Euclidean submanifold $\mathcal{M}_{+}$ onto $\Sigma_{+}$, closing the contour of integration.
}
\label{fig: prepforlorentzian2}
\end{figure}
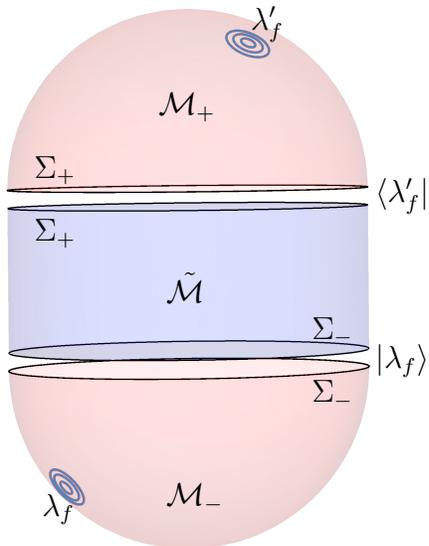

 
Holographically, the preparation of a CFT state via Euclidean path integrals can be mapped to the preparation of a bulk gravitational state on a bulk Cauchy slice $\Sigma_{-}$. Specifically, according to the holographic dictionary, the boundary values of the bulk fields in a southern Euclidean AdS submanifold $\mathcal{M}_{-}$ specify the reference state $|\lambda_i\rangle$ and the sources $\lambda_f$ entering into state preparation. When the fields are on-shell, the boundary values of the fields uniquely determine their values on the Cauchy slice $\Sigma_{-}$, representing the target state $|\lambda_f\rangle$. The time evolution of the CFT state then follows from solving the bulk Einstein's equations with such initial data, which is represented by a section of a Lorentzian cylinder, as depicted in Figure \ref{fig: prepforlorentzian2}. Finally, one can close the contour (\emph{e.g.} to compute transition amplitudes) by gluing  another Euclidean section at the end of the cylinder, representing a path integral preparing the state $\langle \lambda_f'|$.

State preparation offers an intuitive description of field theory complexity, which has features similar to Nielsen's geometric complexity proposal. To define computational complexity, we need to associate a cost to the mapping in (\ref{eq:statelambdatest}).
The precise definition follows from recognizing that the space of coherent states $|\lambda\rangle$  is described by a manifold coordinatized by $\{\lambda_{\alpha}\}$, and is endowed with a symplectic form $\Omega_{\text{bdry}}(\delta_{1}\lambda,\delta_{2}\lambda)$, where $\delta_{1,2}$ refer to arbitrary deformations of the sources. Distances in the space of sources are given in terms of a metric $g_{ab}$, where the minimal path in this space is found by minimizing a `cost' function $F$, represented by the kinetic energy $F=g_{ab}\dot{\lambda}^{a}\dot{\lambda}^{b}$ \cite{Belin:2018bpg}. The computational complexity $\mathcal{C}$  between a given reference state,  defined by some set of sources $\lambda_{i}$, and a target state prepared by sources $\lambda_{f}$ then amounts to identifying a `particle' trajectory which minimizes the kinetic energy. Intuitively, the set of sources $\{\lambda_{f}\}$ act as the set of gates $\{g_{j}\}$ comprising the unitary in (\ref{eq:defcomp}). One may consider variations of the complexity with respect to $\lambda_{f}$, 
which can be used to look for variations which minimize the computational cost, $\delta_{\lambda_{f}}\mathcal{C}=(\dot{\lambda}^{a}|_{\lambda_{f}})g_{ab}\delta\lambda^{b}_{f}$. Thus, complexity obeys a first law \cite{Bernamonti:2019zyy}.  
In particular, consider the special deformation of the sources, denoted by $\delta_{C}\lambda$, such that $g(\delta_{C}\lambda,\delta\lambda)=\Omega_{\text{bdry}}(\delta_{C}\lambda,\lambda)$.
Then, for such deformations
\beq \delta_{\lambda_{f}}\mathcal{C}=\Omega_{\text{bdry}}(\delta_{C}\lambda,\delta\lambda)\;.\label{eq:firstlawcomnograv}\eeq
It is worth emphasizing that this first law is purely a field theory statement. 

We propose that the principle of spacetime complexity is naturally captured by state preparation. Namely, spacetime dynamics emerges from varying complexity.  We provide concrete evidence of this principle by deriving the \emph{linearized} Einstein's equations via the first law (\ref{eq:firstlawcomnograv}), which naturally refers to small variations of sources $\lambda_{f}$, or, equivalently, perturbations of the target state $|\lambda_{f}\rangle$. While we only derive the linearized equations, we expect non-linear contributions to Einstein's equations to similarly arise when we move beyond linear order perturbations of the target state, analogous to the derivation of Einstein's equations from entanglement entropy \cite{Faulkner:2017tkh}.

Let us make this discussion mathematically concrete. The key insight is that  the mapping between boundary sources and initial data extends to the symplectic structure of both the boundary CFT and its bulk gravitational counterpart, such that there is an  equivalence between boundary and bulk symplectic forms \cite{Belin:2018fxe}
\beq \Omega_{\text{bdry}}(\delta_{1}\lambda,\delta_{2}\lambda)=\int_{\Sigma}\omega_{\text{bulk}}(\phi,\delta_{1}\phi,\delta_{2}\phi)=\Omega_{\text{bulk}}(\delta_{1}\phi,\delta_{2}\phi)\,.
\label{eq: boundaryEqualsInitial}\eeq
To arrive at this expression one invokes the extrapolate AdS/CFT dictionary to relate sources $\lambda$ to fields $\phi$ living in the bulk AdS spacetime, including the metric, such that source variations correspond to variations of the bulk fields. The boundary symplectic form $\Omega_{\text{bdry}}$ is proportional to variations of the on-shell bulk gravitational action with respect to the fields, which may be covariantly expressed as an integral of the symplectic current $\omega_{\text{bulk}}$
 over the southern hemisphere of Euclidean AdS. When the  arbitrary field variations $\delta_{1,2}\phi$ obey the \emph{linearized} equations of motion, then $d\omega_{\text{bulk}}=0$ may be `pushed' to an  initial value surface $\Sigma$. Finally, when the bulk fields are deformed by $\delta_{C}$ in (\ref{eq:firstlawcomnograv}), one combines (\ref{eq:firstlawcomnograv}) and (\ref{eq: boundaryEqualsInitial}) to arrive at a first law of holographic complexity  \cite{Belin:2018fxe,Belin:2018bpg}
\beq
\delta\mathcal{C}=\Omega_{\text{bdry}}( \delta_C \lambda, \delta \lambda) = \Omega_{\text{bulk}}( \delta_C\phi, \delta\phi) \;.
\label{eq:bulksymdeltav0}
\eeq

We have arrived at the first law (\ref{eq:bulksymdeltav0}) using a specific definition of complexity given in terms of state preparation, but we expect first laws of complexity to hold more generally \cite{Bernamonti:2019zyy}. In particular, Eq. (\ref{eq:bulksymdeltav0}) is consistent with the `complexity=volume' (CV) conjecture  \cite{Susskind:2014rva,Susskind:2014jwa,Stanford:2014jda,Couch:2016exn}. Precisely, 
the complexity $\mathcal{C}$ of a CFT state defined on a boundary Cauchy slice $\sigma$
is dual to the volume $V$ of a maximal (bulk) hypersurface $\Sigma$ homologous to $\sigma$,
\beq
\mathcal{C}=\frac{V}{G_{N}\ell}\,.\label{eq:CVconj}
\eeq
Here $G_{N}$ is Newton's gravitational constant and $\ell$ is some bulk length scale (such as the curvature scale of AdS).
Note that when CV duality was initially proposed, the precise description of field theory complexity was not particularly well-defined. Rather, ordinary quantum computational complexity was proposed to be a natural information theoretic quantity to describe the late time growth of the wormhole connecting the two sides of an eternal black hole \cite{Susskind:2014moa}. With this in mind, it was shown in \cite{Belin:2018fxe,Belin:2018bpg} there is a particular deformation of bulk fields $\phi$ - the  `new York' transformation $\delta_Y$ - for which a first law is given by the variation of the maximal volume $\delta V$ when $\delta_{Y}$ on-shell \cite{York:1972sj}
\beq
\frac{\delta V}{G_{N}\ell}=\Omega_{\text{bulk}}( \delta_Y\phi, \delta\phi)=\Omega_{\text{bdry}}( \delta_Y \lambda, \delta \lambda)=\delta\mathcal{C}\,.
\label{eq:bulksymdeltav}
\eeq
Note, arriving at \eqref{eq:bulksymdeltav} technically does \emph{not} assume CV duality (\ref{eq:CVconj}), such that $\mathcal{C}$ simply denotes the CFT quantity dual to $V$. As emphasized in \cite{Belin:2018bpg}, proving $\delta_Y \lambda = \delta_C \lambda$  amounts to a proof of the CV conjecture with the definition of complexity given in terms of state preparation. 

The important point is that in all instances variations in complexity are related to the \emph{bulk} symplectic form.  Therefore, by imposing the first law to arbitrary initial data, we may derive the \emph{covariant} linearized Einstein's equations. For illustrative purposes, let us consider perturbations about vacuum AdS, such that the bulk field $\phi$ solely represents the spacetime metric $g_{\mu\nu}$, and where we denote the linearized Einstein's equations by $\delta E^{\mu\nu}=0$. From Stokes' theorem and assuming only the first law (\ref{eq:bulksymdeltav}), we find Einstein's equations must hold in the (Euclidean) section of AdS which prepares the initial state \cite{Pedraza:2021mkh,Pedraza:2021fgp}
\begin{equation}
d \omega_{\text{bulk}}( \delta_Y g_{\mu\nu}, \delta g_{\mu\nu}) = -\delta E^{\mu \nu} \delta_Y g_{\mu \nu}=0\qquad\Rightarrow\qquad \delta E_{\mu\nu}=0\;.
\label{eq:closedness}\end{equation}
 Equation \eqref{eq:closedness} implies initial data prepared on $\Sigma$ is on-shell.
 Demanding that this holds in all Lorentz frames, we conclude the \textit{Lorentzian} Einstein's equations hold in the AdS cylinder. 
 
 While the derivation (\ref{eq:closedness}) assumes a particular form of the first law, we expect Einstein's equations will arise from varying complexity more generally. Indeed, there exist multiple proposals for the holographic dual of CFT state complexity, including a specific on-shell action \cite{Brown:2015bva,Brown:2015lvg}, or possibly `anything' \cite{Belin:2021bga}, reflecting the fact complexity is innately ambiguous. 
 For example, there is an ambiguity in choosing the gate set with which to transform a reference into its target state, or using a different cost function. Consequently, the gravitational dual for complexity should reflect these ambiguities. With respect to holographic state preparation, the complexity $\mathcal{C}$ obeys the first law (\ref{eq:bulksymdeltav0}) when sources are deformed via an on-shell perturbation $\delta_{C}$,
 not necessarily the new-York deformation, and Einstein's equations will similarly arise from varying complexity.

Hence, assuming holographic duality, the first law of complexity implies the linearized Einstein's equations around vacuum AdS, or more generally, a reference background (\emph{e.g.} if we allow $|\lambda_i\rangle\neq|0\rangle$). This explicitly captures the spirit of spacetime complexity: optimal quantum computation imposes gravitational field equations. The covariant derivation (\ref{eq:closedness}) reflects and extends previous work \cite{Czech:2017ryf,Caputa:2018kdj,Susskind:2019ddc}, which offered preliminary hints on the connection between optimal computation and the laws of gravity.



Returning to CV duality (\ref{eq:CVconj}), it is appealing in that we are reminded of another connection between information theory and spacetime geometry: the Ryu-Takayanagi (RT) entropy-area prescription \cite{Ryu:2006bv,Hubeny:2007xt}. The relation, a generalization of the Bekenstein-Hawking entropy-area formula for black holes, proposes the area $\mathcal{A}$ of a bulk minimal surface anchored to the boundary of AdS is equal to the entanglement entropy $S(A)$ of a CFT state restricted to a subregion $A$ homologous to the minimal surface
\beq S(A)=\frac{\mathcal{A}}{4G_{N}}\;.\label{eq:RTform}\eeq
 
\begin{figure}[t]
\captionsetup{singlelinecheck = false, justification=justified}
{\centering
$\quad$\includegraphics[width=2.1in,trim=0 -0.2cm 0 0]{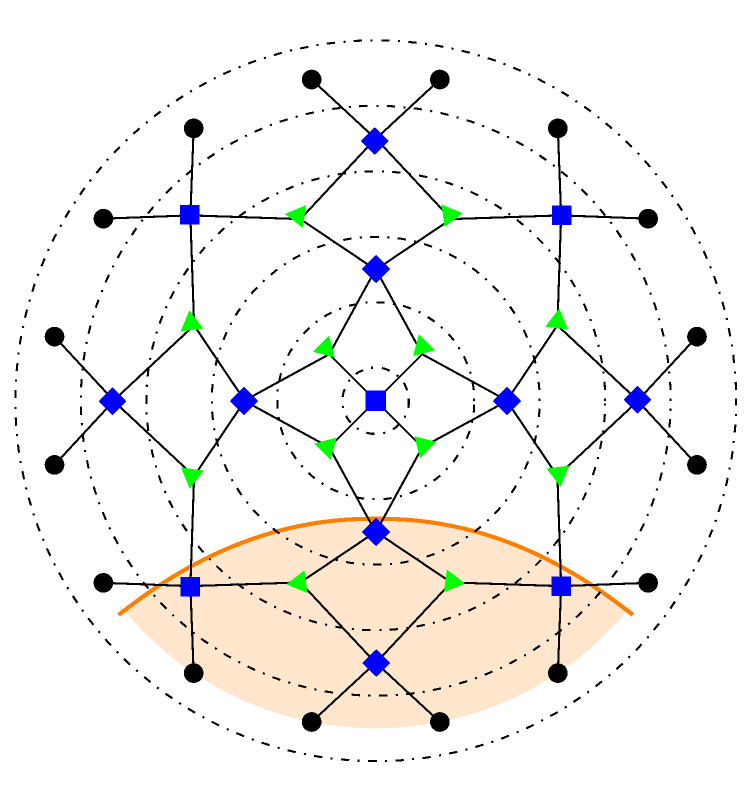}$\quad\quad$ \includegraphics[width=3.1in,trim=-1cm 2.1cm 0 0]{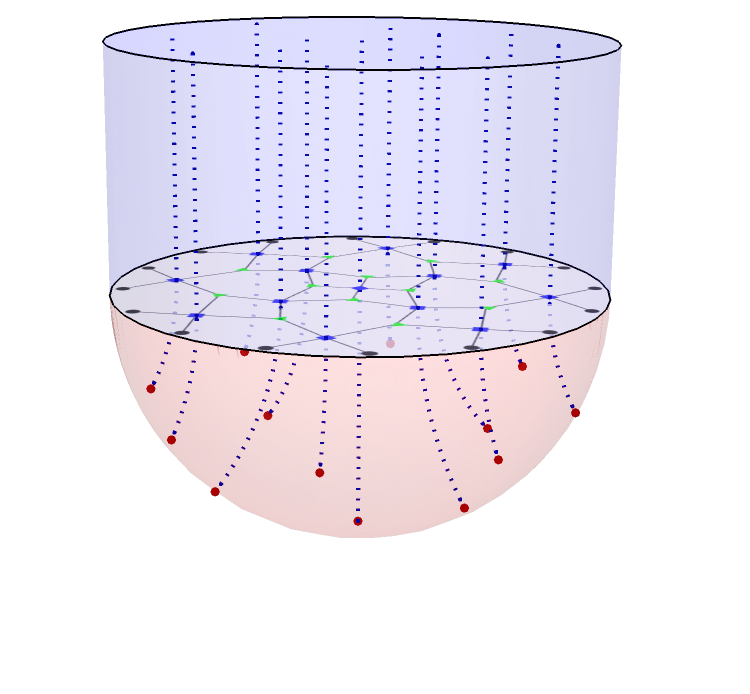}
 \begin{picture}(0,0)
\put(-337,-2){$A$}
\put(-120,110){$v$}
\put(-41,71){$\Sigma$}
\put(-110,23){$\mathcal{M}_{-}$}
\put(-57,98){$\tilde{\mathcal{M}}$}
\end{picture}}
\caption{\small Left: Network discretization of CFT state. For a subregion $A$, entanglement entropy is computed as the the minimal number of cuts through the network. Right: Complexity is equal to the minimum number of gatelines preparing a state on the maximal volume slice $\Sigma$. Each gateline sourced from the boundary attaches a unitary gate to each tensor in a tensor network discretization of $\Sigma$.
Together, spacetime is a collection of tensor networks connected via Lorentzian flows $v$.
\label{fig:TNs}}
\end{figure}
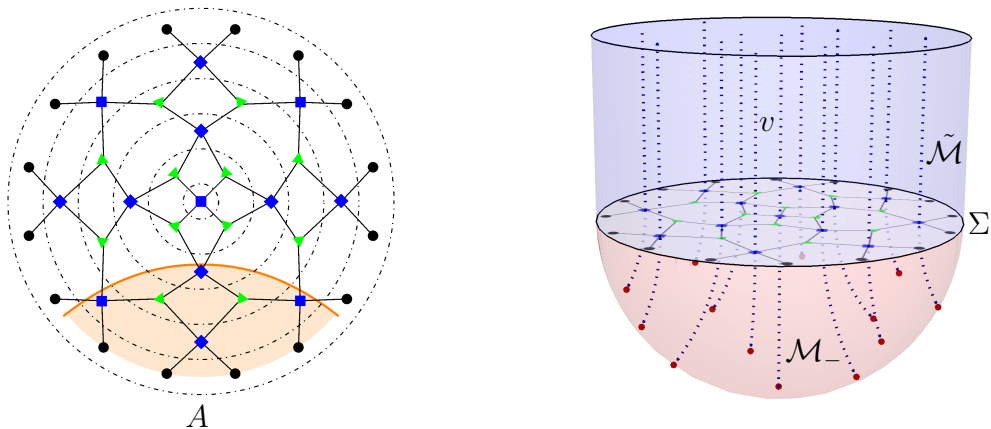

A salient feature of the RT prescription is the `entanglement=geometry'  paradigm \cite{VanRaamsdonk:2010pw,Bianchi:2012ev,Balasubramanian:2014sra}: spatial connectivity is generated by entanglement. This is beautifully captured by tensor network models of AdS/CFT \cite{Vidal:2007hda,Vidal:2008zz,Swingle:2012wq}. Tensor networks represent spatial discretizations of quantum states, where links between tensors in the network represent spatial correlations between different degrees of freedom. Remarkably, particular networks that discretize quantum critical states yield an emergent AdS metric, geometrizing spatial correlations across energy scales.
Further, in this context, entanglement entropies are computed by cutting links in the network, consistent with the RT prescription  (\ref{eq:RTform}), such that $S(A)$ is equal to the minimum number of cuts (see Figure \ref{fig:TNs}). Altogether, tensor networks neatly illustrate the connection between entanglement and the emergence of space. 

A more precise connection between holographic entanglement and tensor networks is given by the `bit thread' reformulation of the RT formula (\ref{eq:RTform}) \cite{Freedman:2016zud,Headrick:2017ucz,Agon:2018lwq}.  In this context the minimal area surface calculating the entanglement entropy $S(A)$ is replaced by the maximum flux of a divergenceless, Riemannian vector field $v$ through $A$,
\beq
S(A)=\underset{v}{\text{max}}\int_{A}\hspace{-1mm}v\;,
\label{eq:RTasflowint}
\eeq
such that area minimization is mapped to flux maximization. The equivalence between the two prescriptions (\ref{eq:RTform}) and (\ref{eq:RTasflowint}) follows from an application of the continuous version of the max flow-min cut theorem, a well-known principle in network theory.

Tensor network models also provide a natural realization of CV duality (\ref{eq:CVconj}) \cite{Stanford:2014jda}. One may associate a fixed spatial volume to each physical tensor, such that the complexity of the discretized state is equal to the minimum number of tensors necessary to describe the network. Thus, while entanglement builds space \cite{VanRaamsdonk:2010pw}, complexity quantifies the amount of space being built. When we combine this viewpoint with state preparation, we are led to a sharp picture of spacetime complexity. 

To appreciate this, we first reformulate CV duality (\ref{eq:CVconj}) using a continuous version of the min flow-max cut theorem \cite{Pedraza:2021mkh,Pedraza:2021fgp}
\beq
\!\mathcal{C}=\underset{v}{\text{min}}\int_{\partial\mathcal{M}_{-}}\hspace{-3mm} v\;,
\label{eq:cvreformin}
\eeq
where we have replaced the maximization of volume with the minimization of the flux of a timelike, divergenceless vector field $v$ -- a Lorentzian flow \cite{Headrick:2017ucz}. 
We can understand our reformulation as the Lorentzian analog of (\ref{eq:RTasflowint}), where the holographic complexity $\mathcal{C}$ is equal to the minimum number of Lorentzian `threads' 
passing through $\Sigma$.
Importantly, the continuous min flow-max cut theorem requires the entire manifold be compact, hence, for (\ref{eq:cvreformin}) to hold, we must attach Euclidean portions to the Lorentzian cylinder, naturally connecting to the prescription of state preparation. Conceptually, this suggests we should understand threads  as preparing the target state on $\Sigma$ from the reference state defined on the lower hemisphere $\partial\mathcal{M}_{-}$, as visualized in Figure \ref{fig:TNs}.


More accurately, threads 
enter from the (southern) Euclidean submanifold, each attached to boundary sources $\lambda_{f}$, and pass through $\Sigma$ (see Figure \ref{fig:TNs}). A minimal flux configuration is then one which
optimally prepares the CFT state on $\Sigma$, namely, the configuration requiring fewer operations to assemble the state. In this way, Lorentzian threads act as \emph{gatelines}: timelike trajectories representing  unitary gates needed to transform a reference state $|\lambda_{i}\rangle$ into a target state $|\lambda_{f}\rangle$. Complexity, therefore, is the minimum number of gatelines through $\Sigma$ preparing the target state.
A new and notable feature of this gateline picture is that reference states play a crucial role, unlike the original CV proposal, where, mysteriously, the reference state need not be defined. In particular, the complexity of the vacuum is dependent on the choice of reference state. If, for example, the reference state is the CFT vacuum $|0\rangle$ (where all sources $\lambda_{i}$ are turned off), the complexity vanishes since the southern hemisphere shrinks to a point, and the volume goes to zero.

The gateline interpretation deepens our insight into tensor network constructions of spacetimes. Specifically, an optimal thread configuration $v$ prepares the tensor network on $\Sigma$. We can imagine attaching a unitary to each thread, connecting to each  physical tensor of the network. These unitaries 
 transform a reference state to its target. Upon analytic continuation, this operation generates time evolution
and the network acts as a quantum circuit. From this follows an apt visualization: Lorentzian threads sew together tensor networks discretizing slices foliating spacetime, offering an emergent notion of time and leading to a literal construction of the fabric of spacetime \cite{Pedraza:2021mkh,Pedraza:2021fgp}.

Therefore, when Lorentzian flows are taken to be fundamental, spacetime geometry is a derived concept. Even further, the flow picture is consistent with the fact spacetime dynamics is naturally captured by varying complexity \cite{Pedraza:2021mkh,Pedraza:2021fgp}. This follows from the observation that a natural choice for an optimal  Lorentzian thread configuration characterizing perturbations is the symplectic current $\omega_{\text{bulk}}(\delta_{Y},\delta)$, where the divergenceless condition demands $d\omega_{\text{bulk}}=0$, thereby imposing the linearized Einstein's field equations (\ref{eq:closedness}).

We have explicitly realized spacetime complexity at the linearized level in the context of the AdS/CFT correspondence, however, we expect the principle holds more generally. In part, this is because whatever quantum gravity may ultimately be, it arguably has an information theoretic origin, for which complexity plays a prominent role.
More fundamentally, computation is universal.
A similar observation was made about thermodynamics, in which Einstein's equations arise from a locally holographic  implementation of the Clausius relation \cite{Jacobson:1995ab}. Computation, however, is more essential than coarse grained equilibrium thermodynamics, and, echoing sentiments of Lloyd \cite{Lloyd:2005js,Lloyd:2013xba}, it is tantalizing to imagine a unifying framework where fundamental physics may be treated as a quantum computer.





\noindent \emph{Acknowledgements.}
It is a pleasure to thank  C\'esar Ag\'on, Jos\'e Barb\'on, Elena C\'aceres, Roberto Emparan, Willy Fischler, Matthew Headrick, Michal Heller, Ted Jacobson, Rob Myers, Maulik Parikh, Martin Sasieta, Leonard Susskind, Tadashi Takayanagi, Marija Tomašević, Manus Visser and George Zahariade for discussions and useful correspondence. JFP is supported by the `Atracci\'on de Talento' program (2020-T1/TIC-20495, Comunidad de Madrid) and by the Spanish Research Agency (Agencia Estatal de Investigaci\'on) through the Grant IFT Centro de Excelencia Severo Ochoa No CEX2020-001007-S, funded by MCIN/AEI/10.13039/501100011033. AR and AS are supported by the Simons Foundation via \emph{It from Qubit: Simons Collaboration on quantum fields, gravity, and information}, and by EPSRC. ZWD is supported by Perimeter Institute; research at Perimeter Institute is funded by the Government of Canada through the Department of Innovation, Science and Economic Development Canada and by the Province of Ontario through the Ministry of Economic Development, Job Creation and Trade.

\bibliography{refs-CBT}

\end{document}